\documentclass[amsmath,showpacs,twocolumn,aps,prl]{revtex4}
\usepackage{bm}
\usepackage{graphics}
\begin{document}
%\title{Intersubband Edge Singularity in Metallic Nanotubes}

\title{Breakdown of classical electrostatics in the depolarization of quantum wires and nanotubes}

\author{L.~Shan}
\affiliation{Department of Physics and Astronomy, University of Utah, Salt Lake
City, Utah 84112, USA}

\author{E.~G.~Mishchenko}
\affiliation{Department of Physics and Astronomy, University of Utah, Salt Lake
City, Utah 84112, USA}

\begin{abstract}
In quantum wires, such as metallic nanotubes,  the optical absorption of the transverse polarization is controlled by the depolarization effect which stems from the redistribution of conduction electrons around the circumference of the system. The traditional electrostatics treatment of the depolarization effect relies on approximating the system by a cylinder with some effective dielectric permittivity. We demonstrate that this simple intuitive picture does not adequately describe optical absorption near its threshold, as  the depolarization effect becomes dominated by many-body correlations which strongly modify the  spectral dependence of absorption.
\end{abstract}

\maketitle

{\it Introduction}. Quantum wires have a limited number of conducting channels $ N$, which originate from confinement of electrons in the transverse direction \cite{DBG}. Transitions between such channels determine the response of a wire to a transverse electric field. For example, in the case of a carbon nanotube \cite{SDD}, or similarly rolled two-dimensional hexagonal crystals, such as tungsten sulfate and gallium nitride nanotubes, the transverse polarization induced by the electric field arises from the motion of carriers around the circumference of the nanotube. In the quantum picture, this motion amounts to a redistribution of electrons between different subbands corresponding to different values of the azimuthal angular momentum $m$. According to the band structure of metallic carbon nanotubes (see Fig.~1), the subbands with $m\ne 0$ are separated by an energy  gap $\Delta$ from the gapless right- and left-moving states with $m=0$, the latter giving rise to longitudinal dc conduction of nanotubes.

The field-driven redistribution of electrons reduces the value of the electric field inside the wire. This phenomenon, known as the depolarization effect, has first been addressed in the context of nanotubes in Ref.~[\onlinecite{AA}] and further studied in Refs.~[\onlinecite{BLC,TMY,MRV}].  Because of the limited number of transverse channels, the  suppression is not complete. In the minimal electrostatic model, one can treat the wire as a solid cylinder with some effective dielectric permittivity $\varepsilon_\perp$. From elementary electrostatics \cite{LL} it then follows that the  electric field inside the wire is uniform, $E_{i}=2E_0/(1+\varepsilon_\perp)$, and reduced compared with the applied external field $E_0$. In terms of the effective transverse susceptibility $\alpha_\perp = (\varepsilon_\perp -1)/4\pi$, this can be restated in the equivalent form,
\begin{align}
\label{electrostatics RPA series}
E_i = \frac{E_0}{1+2\pi \alpha_\perp} = E_0\left[ 1-2\pi\alpha_\perp +(2\pi\alpha_\perp )^2 - ...\right].
\end{align}

The  meaning of the consecutive terms in this geometric series is rather transparent: The external field $E_0$ induces the surface charge density $\sigma = \alpha_\perp E_0 \cos\theta$, where $\theta$ is the circumferential angle; this charge density produces the correction $E^{(1)}=-2\pi \alpha_\perp E_0$ to the field $E_0$. In turn, this correction induces additional charge density, and so on, resulting in the infinite series, Eq.~(\ref{electrostatics RPA series}).

The same approach can be extended to an ac field \cite{AA,BLC} as long as its frequency $\omega$ is low enough for the retardation to be disregarded:  $\omega \ll c/R$, where $R$ is the radius of the nanotube. Such condition is always satisfied for optical frequencies. Above the threshold frequency $\omega=\Delta$, the transitions between $m=0$ and $m=\pm 1$ subbands become possible (Fig.~1). The imaginary part of $\alpha_\perp(\omega)$ determines the absorption spectrum. Calculating the amount of Joule heat generated per unit length of the tube, we can write it in either of the two equivalent forms,
\begin{equation}
\label{Joule electrostatics}
Q=\frac{\pi }{2} \omega R^2 |E_i |^2 \alpha''_\perp(\omega) =\frac{\pi\omega }{2} R^2 |E_{0}|^2 %\text{Im}
\left(\frac{\alpha_\perp(\omega)}{1+2\pi \alpha_\perp(\omega)}\right)''.
\end{equation}
\begin{figure}
\resizebox{.450\textwidth}{!}{\includegraphics{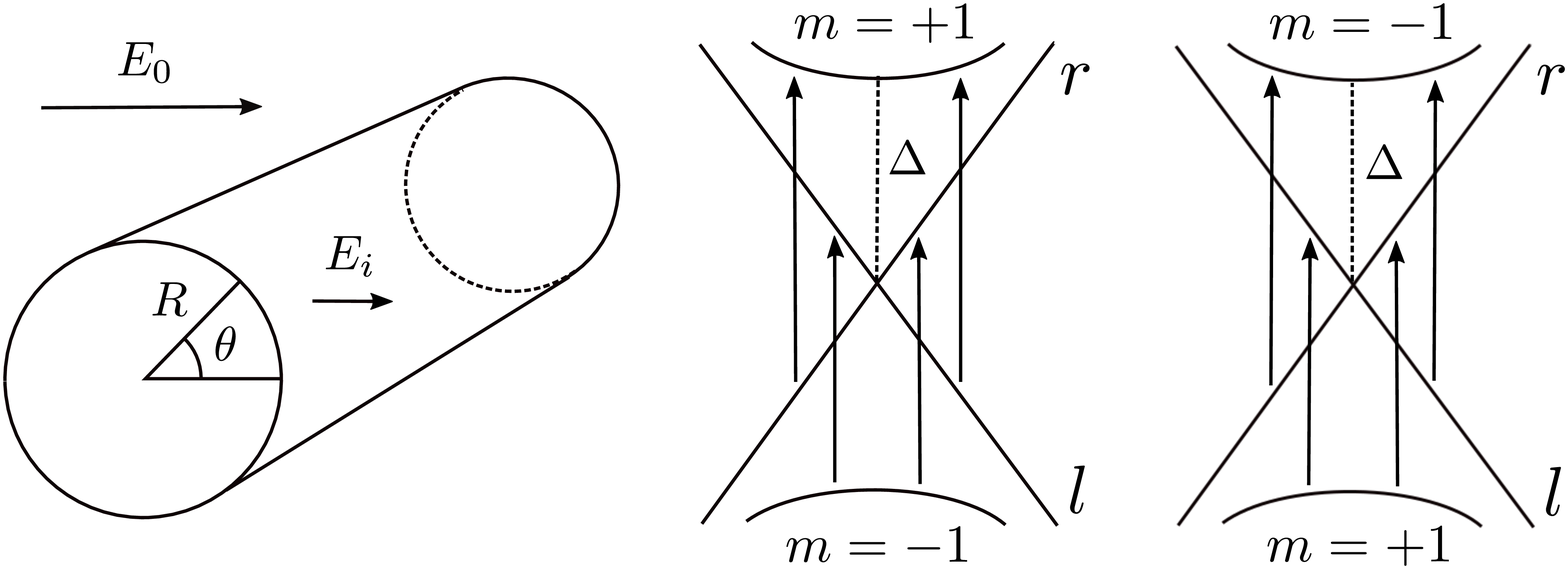}}
\caption{\label{fig1} Geometry of a nanutube in a transverse electric field (left) and the absorption  transitions induced by such field: with the increase of the azimuthal angular momentum  $ m$,
$0 \to +1$ and $-1 \to  0$ (center), and with the decrease of it, $0 \to -1$ and $+1 \to  0$ (right).}
\end{figure}

It is our main finding that such an intuitively appealing electrostatics approach nonetheless breaks down in the most interesting situation -- near the absorption threshold, $\omega\approx \Delta$. Before explaining why it fails to account properly for the electron-electron Coulomb interaction, let us point out that Eqs.~(\ref{electrostatics RPA series}) and (\ref{Joule electrostatics}) are in fact equivalent to the random phase approximation (RPA) \cite{Mahan}. In the latter, a system responds to a Fourier harmonic $\varphi_m (\omega) e^{im\theta}$ of a weak electric potential with  the variation of the particle density,
\begin{equation}
\label{charge density displacement general}
\langle \rho_{m}(\omega) \rangle = \chi_m(\omega) e \varphi_m (\omega),
\end{equation}
whose magnitude is determined by the polarization function
\begin{equation}
\label{correlation function Kubo}
\chi_m(\omega) = -i\int\limits_0^{\infty}dt~e^{i\omega t}\int\limits_{-\infty}^{\infty}dx~\Bigl<\Bigl[\hat \rho_m(x,t),\hat \rho_{-m}(0,0)  \Bigr]\Bigr> ,
\end{equation}
which is a correlator of the electron density operators $\hat \rho_{m}(x,t)$ expressed as a sum over the channels  with different azimuthal angular momenta: $\hat \rho_{m}(x,t)=\sum_\mu~\hat \psi_\mu^\dagger(x,t)\hat \psi_{\mu+m}(x,t)$.

Because the potential, $\varphi =-RE_0(t)\cos\theta$,  of a homogeneous  external electric field $E_0$  contains only the dipolar Fourier harmonics,  $\varphi_{\pm 1} = -E_0R/2$, only the $m=\pm 1$ components of the charge density can be excited. From symmetry it follows that $\langle \rho_{1} \rangle=\langle \rho_{-1} \rangle$. Utilizing  Eq.~(\ref{charge density displacement general}) to relate the potential to the surface charge  density, $\sigma =  e \langle \rho_{1} \rangle \cos\theta/(\pi R) $, and then using the charge conservation law  to find the sheet electric current flowing around the surface of the wire, one can calculate the amount of Joule heat dissipated in the wire,
\begin{equation}
\label{Joule heat}
Q=-\frac{\omega}{4} R^2 e^2|E_{0}|^2 \chi''(\omega).
\end{equation}
[To ease notations, we  omit the subscript $1$ in $\chi_1(\omega)$.]

When Coulomb interaction is ignored, the correlation function (\ref{correlation function Kubo}) follows from the elementary Lindhard calculation \cite{AA} for the band structure of a nanotube, obtained by ``rolling'' a $\pi$-band of graphene \cite{SDD}. At zero temperature, the only transitions with the change of the angular momentum $+1$ that are possible are those between the gapless right-movers $\epsilon =vp$ or left-movers $\epsilon =-vp$ subbands with $m=0$ and the first gapped subbands $|\epsilon| = \sqrt{\Delta^2+v^2 p^2}$,  where $\Delta =v/R$ is the energy gap; see Fig.~1. All four such processes contribute equally. For positive $\omega$:
\begin{eqnarray}
\label{chi non-interacting}
\chi^{(0)}(\omega) &=& \frac{N}{\pi} \int\limits_0^\infty  \frac{dp}{\omega -vp -\Delta - p^2/2m^*+i\eta} \nonumber\\
 &\approx & \frac{N}{v\pi}\Bigl[\ln{|\Omega|} - i\pi \Theta (-\Omega)\Bigr], \hspace{0.4cm} \Omega= \frac{\Delta-\omega}{\Delta},
\end{eqnarray}
Absorption can only occur if the frequency of the external field exceeds the gap, $\omega>\Delta$.
In nanotubes, there are two orbital valleys and two spin directions, so that the total electron flavor degeneracy $N=4$.

Consider now the Coulomb interaction of electrons,
\begin{equation}
\label{Coulomb_interaction}
\hat V=\frac{1}{2}\sum_{ \mu \nu m} V_m \int%\limits_{-\infty}^{\infty}
dx \,  \hat \psi_{\mu +m}^\dagger (x) \hat  \psi_{\nu-m}^\dagger (x) \hat \psi_{\nu}(x) \hat \psi_{\mu}(x),
\end{equation}
with the matrix elements $V_m$ for the scattering events that occur with the change of the angular momentum $m$ previously found in Ref.~[\onlinecite{MAG}].  For non-zero $m$, $V_m={e^2}/{|m|}$, while $V_0$ happens to be  logarithmically stronger. For example, $V_0 = 2e^2 \ln{(d/R)}$, in a setting where a metallic gate is  located some distance $d$ away.

Coulomb interaction strongly modifies the absorption lineshape even though traditional exciton bound states  \cite{ANS} cannot be formed. Indeed, the absence of backscattering forbids velocity reversal of the gapless states. As a result, attractive Coulomb interaction between electrons and holes cannot confine them to a finite region of space.

Suppose first that one ignores $V_0$  and only retains $V_1=e^2$ to account for the transverse polarization of the system. Because each closed electron loop brings  the flavor degeneracy $N$, in the formal limit of  $N\gg 1$ one can retain only the diagrams with the maximum possible number of the loops. This yields the RPA geometric series for  the density-density correlation function (\ref{correlation function Kubo}), which consists of consecutive loops connected by the interactions $V_1$, as shown in Fig.~2,
\begin{equation}
\label{chi_RPA}
RPA:\hspace{0.4cm} \chi (\omega) = \frac{\chi^{(0)}(\omega)}{1-V_1\chi^{(0)}(\omega)}.
\end{equation}
This expression together with Eq.~(\ref{Joule heat}) reproduces the dissipated power (\ref{Joule electrostatics}) as long as the transverse susceptibility is identified with the {\it non-interacting} ($V_0=0$) polarization operator:  $\alpha_\perp(\omega) = -\frac{e^2}{2\pi}\chi^{(0)}(\omega)$. This result predicts that the threshold absorption at $\Omega=0$, rather than being a step function,  as in $\text{Im}\chi^{(0)}(\omega)$, is suppressed as $\ln^{-2}\Omega $.

Ignoring $V_0$, however,  can be problematic in one-dimensional systems \cite{LM,DL,H}. For example, in nanotubes, the density of states of the gapless subbands has a power-law suppression, $\nu_0 (\epsilon) \propto
\epsilon^\alpha$  \cite{KBF,McEuen1,Dekker,Ishii}. The exponent $\alpha=(1-g)^2/2Ng$ is customarily expressed via  the effective coupling constant
$g=v/u$, the ratio of the band velocity $v$ and the velocity of the collective charged
plasmon modes, $u=v\sqrt{1+NV_0/\pi v}$. Similarly, the density of gapped states is suppressed as well, $\nu_1
(\epsilon) \sim (\epsilon -\Delta)^{-1/2+\beta}$, compared with the non-interacting case \cite{B}; the suppressing exponent is
$\beta=(1-g^2)^2/2Ng$. These
non-perturbative renormalizations originate from the decomposition of a single-electron state into an infinite number of charged plasmon modes \cite{G}.

The modification of the polarization function (\ref{chi non-interacting}) by the $V_0$ interaction was studied in Ref.~[\onlinecite{MS}]. In contrast to the density of states, the polarization function is {\it enhanced} at low $\Omega$:
\begin{equation}
\label{polarization operator MS}
\chi ^{(0)}_{_V}(\omega) = -\frac{N\Gamma(\gamma)}{v\pi}\Bigl[\Omega^{-\gamma} -1\Bigr], \hspace{0.5cm} \gamma=\frac{2-g-g^3}{2N},
\end{equation}
with the subscript in $\chi ^{(0)}_{_V}$ indicating that the interaction $V_0$ is taken into account (the superscript reminding that one still assumes $V_1$=0). It is worth noting that the gapped state created in the interband transition acts similarly to a core-hole in the conventional x-ray edge singularity problem studied by Mahan \cite{Mahan2} and Nozi\`{e}res and De Dominicis \cite{NDD}. The one notable difference is that in addition to the interaction with a core-hole, electrons themselves are strongly interacting. Nonetheless, the edge enhancement in Eq.~(\ref{polarization operator MS}) is the result of a competition between the Mahan-type enhancement and the Nozi\`{e}res-De Dominicis orthogonality catastrophe-type suppression for a given value of the interaction strength $g$.

The polarization function (\ref{polarization operator MS}) determines the response of the nanotube to the actual electric field $E_i$ inside  it, $Q \propto  -\text{Im} \chi_V^{(0)}|E_i|^2 \propto \Omega^{-\gamma} |E_i|^2$; cf.~the first identity in Eq.~(\ref{Joule electrostatics}). However, to  relate the inside field $E_i$ to the applied field $E_0$ one also has to take into account  the dipolar interaction $V_1$. It is the latter that is responsible for the redistribution of electrons around the circumference of the wire in response to a perpendicular electric field.

If one used the electrostatics/RPA mean-field approach embodied in Eqs.~(\ref{Joule electrostatics}) and (\ref{chi_RPA}),  one would arrive at the absorption, $Q \propto \Omega^\gamma |E_0|^2$,  {\it suppressed} with the same  exponent  $\gamma$.

Below we show that the depolarization suppression of the near-threshold absorption is in fact much stronger,
\begin{equation}
Q\propto  (\omega-\Delta)^{3\gamma} |E_0|^2,
\end{equation}
and demonstrate that the  conventional electrostatics fails because the independent-loop approximation does not apply in the presence of $V_0$-correlations between different loops, as illustrated in Fig.~2.
\begin{figure}
\resizebox{.30\textwidth}{!}{\includegraphics{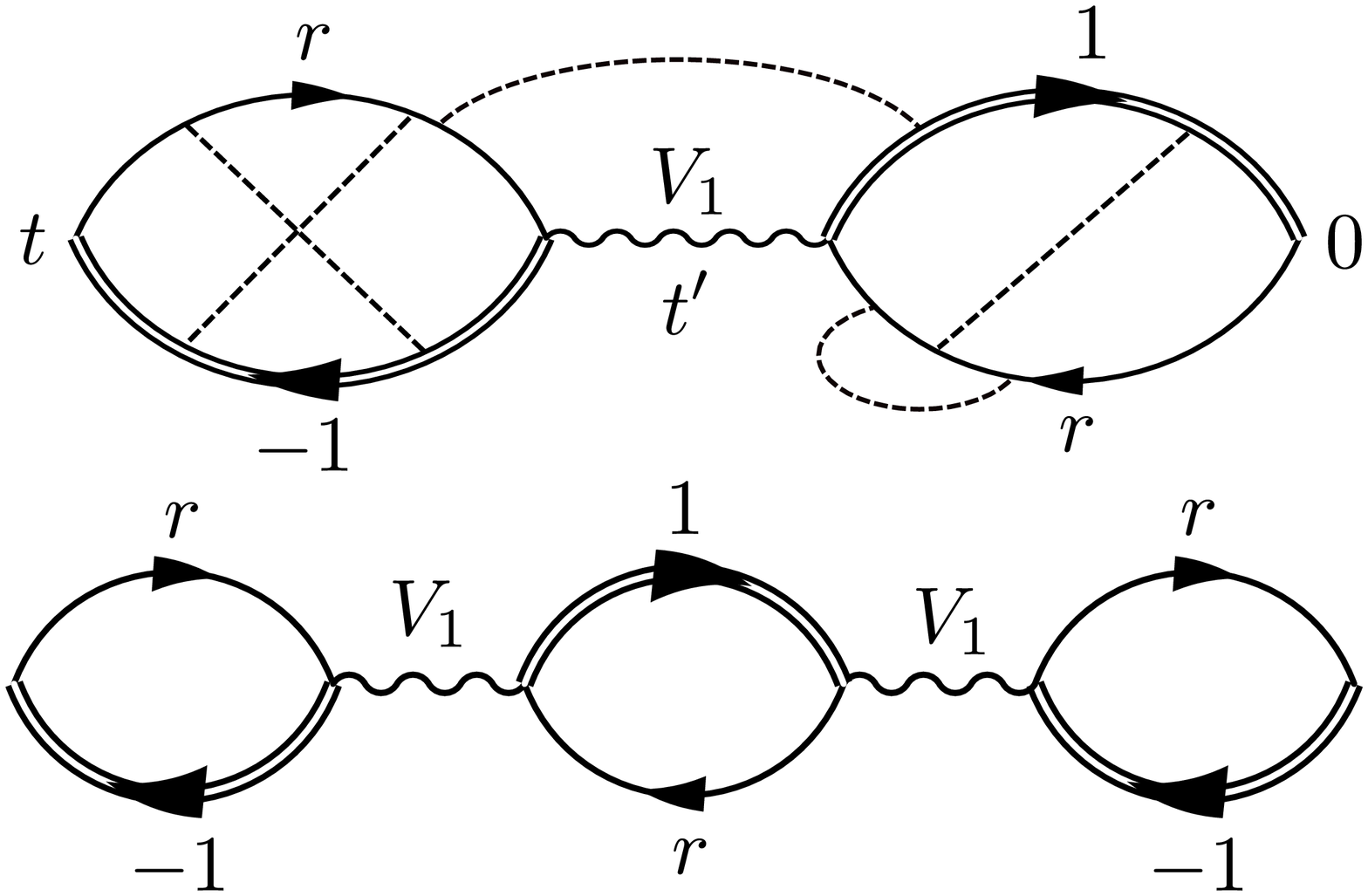}}
\caption{\label{fig2} First- and second-order corrections to the polarization function $\chi_1(\omega)$ in the interaction $V_1$ (wavy line). The interaction $V_0$ (dashed line) is    accounted to all orders, including interactions between different loops (not shown in the second-order diagram). The diagrams providing the leading contributions are shown. Massive gapped electrons $m=\pm 1$ are indicated with a double line; gapless states with a single line. }
\end{figure}

To account for the leading interaction $V_0$ to {\it all orders}, we use the bosonization approach \cite{G} wherein the right- and left-moving gapless electronic operators are represented as the exponentials,
\begin{equation}
\label{massless_operator}
\hat \psi_{r,l}(t,x)=\frac{\hat U_{r,l}}{\sqrt{2\pi R}}  e^{\hat k_{r,l}(t,x)},
\end{equation}
of the bosonic phases $\hat k_{r,l}$ with  $\hat U_{r,l}$ being the fermionic counting operators. The inverse tube radius  $1/R$ is chosen as the ultraviolet momentum cut-off. The bosonic phase operators are given by,
\begin{align}
\label{massless_phase}
\hat k_{r,l} (t,x)=\sqrt{\pi}\sum_{q}\frac{1\pm g~ \text{sgn}~
q}{\sqrt{2gN|q|L}}\Bigl[ \hat a_q e^{-i|q|ut+iqx} -c.c. \Bigl]
\nonumber\\ + \sqrt{2\pi}\sum_{i=1}^{N-1}\sum_{q}\frac{\Theta(\pm
q)}{\sqrt{N|q|L}}\Bigl[ \hat b_{iq} e^{-i|q|vt+iqx} -c.c. \Bigl],
\end{align}
with the upper/lower signs corresponding to the right/left-moving electrons, respectively.
The operators $\hat a_q$ represent the fast charged plasmon modes
of the system, while the remaining $N-1$ modes $\hat{b}_{i q}$ are
neutral (arising from the spin and band degeneracy) and propagating with the band velocity $v$; the Hamiltonian of the interacting gapless electrons is thus simply,
$\hat H =u\sum_q |q| \hat a_q^\dagger \hat a_q +\sum_{i=1}^{N-1}v\sum_q |q|
\hat b_{iq}^\dagger \hat b_{iq}$.

Although one cannot fully bosonize massive gapped states, it is possible to take advantage of the fact that for optical transitions near the threshold  the momenta of the massive states are small and the electrons are virtually stationary there, $p/m \ll v,u$ \cite{Fur}. Accordingly, massive states can be represented as a product \cite{B,MS},
\begin{equation}
\label{massive_operator}
\hat \psi_1(t,x)=\hat \psi_1^{(0)}(t,x) e^{\hat K (t,x)},
\end{equation}
 of a free fermion operator $\psi_1^{(0)}$ and the exponential of a phase, which is a time integral of the electric potential of the fluctuating plasmon field:
\begin{equation}
\label{massive_phase}
\hat {K} (t,x)=(1-g^2)\sum_q \sqrt{\frac{\pi }{2Ng|q|L}}\Bigl( \hat a_q e^{-i|q|ut+iqx} -c.c \Bigl) .
\end{equation}

Unlike the symmetric part of the interaction $V_0$,  the dipolar interaction $V_1$ is not readily amenable to the bosonization technique. We are thus going to use a ``hybrid'' approach where $V_1$ is treated by means of the usual diagrammatic technique. For example, the main contribution to the first order in $V_1$ (shown in Fig.~2) turns out to be
\begin{widetext}
\begin{eqnarray}
\label{first order with interaction}
\chi^{(1)}_{_V}(\omega) &=& -\frac{V_1N(N-1)}{4\pi^2 R^2}\int\limits_0^{\infty}dt~e^{i(\omega-\Delta) t} \int\limits_0^{t}dt' \left< e^{-k_r(t)}e^{K(t)}e^{-2K(t')}e^{2k_r(t')}e^{K(0)}e^{-k_r(0)} \right>,
\end{eqnarray}
\end{widetext}
where the averaging is performed over the Gaussian fluctuations of the  bosonic fields. The integrals over the coordinates have already been eliminated from Eq.~(\ref{first order with interaction}) owing to the sharply spatially localized propagators composed of the (slow) free fermion operators $\psi_1^{(0)}$. As a result, all the bosonic fields can be taken at $x=0$ with only their time arguments shown in Eq.~(\ref{first order with interaction}). The composition of the bosonic average in Eq.~(\ref{first order with interaction}), as well as the signs of the operators in the exponents, can be traced to the nature and direction of the electron propagators shown in Fig.~\ref{fig2}.

While the  four different combinations of massive and massless propagators contribute {\it  equally} to $\chi^{(1)}$ when $V_0$ is ignored,  this is no longer the case for $V_0\ne 0$. For example, the first order contribution is different where the electrons of the same chirality (left- or right-movers) propagate around both loops. The most singular contribution comes from the top diagram of Fig.~\ref{fig2}, where both chiralities are the same and which yields the highest negative power, $\propto \Omega^{-3\gamma}$. Indeed, calculating the average in Eq.~(\ref{first order with interaction}) and imposing a small-time cut-off $1/\Delta$ corresponding to high energies, we arrive at,
\begin{align}
\label{chi second order}
\chi^{(1)}(\omega) & =\frac{V_1N^2}{4\pi^2 v^2}\int\limits_{1/\Delta}^{\infty}\frac{dt}{t^\gamma}\,e^{i(\omega-\Delta) t} \int\limits_{1/\Delta}^{t-1/\Delta}
\frac{dt'}{[(t-t')t']^{1-2\gamma}}\nonumber\\ &=
\frac{V_1N^2}{4\pi^2 v^2\gamma^2} \Bigl[\frac{1}{3}\Bigl( \Omega^{-3\gamma} -1\Bigr) -\Bigl(\Omega^{-\gamma} -1\Bigr) \Bigr]
\end{align}
Here the integral is taken in the approximation of small $\gamma$, which is formally realized in the limit of a large number of channels $N\gg 1$. When $\gamma\ll1$, it is the regions near the integration limits that contribute the  most to the integrals.

The leading term in Eq.~(\ref{chi second order}) has a greater power, $\Omega^{-3\gamma}$,  than what could be na\"ively expected ($ \Omega^{-2\gamma}$)  from a simple RPA product of the two loops each given by Eq.~(\ref{polarization operator MS}).
Similarly, we identify the leading contributions to every order in $V_1$ as those that i) contain the same chirality ($r$ or $l$) {\it in all loops} and ii) have  two massive particles either  created or destroyed at each $V_1$ interaction line. In the $n$-th order the leading contribution happens to be $\propto V_1^n\Omega^{-(2n+1)\gamma}$, which can be quickly established from power counting. The actual integrals for the numerical coefficients become very complicated in higher orders.  However, it is possible to uniquely construct a series that obeys the following two properties: The $n$-th order term is an odd-power polynomial in $\Omega^{-\gamma}$, and the coefficients in the polynomial are chosen in such a way that in the limit of $\gamma \to 0$ the $n$-th order term reproduces $\ln^{n+1}{\Omega}$, the correct non-interacting ($V_0=0$)  limit. This leads to the following expression
\begin{equation}
\label{summable series}
\chi^{(n)}_{_V}(\omega) =-\frac{N}{\pi v } \frac{(-\lambda_1)^n}{\gamma^{n+1}}  \sum_{j=0}^n A_j^{(n)} \Bigl(\Omega^{-(2j+1)\gamma }-1\Bigr),
\end{equation}
where the effective dimensionless dipolar coupling constant is $\lambda_1=NV_1/(4\pi v)$; the extra factor $1/4$ originates from the fact that only one out of four possible loops has the strongest singularity at $\Omega=0$.

The coefficients of the leading $j=n$ terms  are $A_n^{(n)} = (n+1)/[2^n(2n+1)]$. The summation of these leading terms can be performed exactly with the identity,
\begin{equation}
\sum_{n=0}^\infty \frac{n+1}{n+\frac{1}{2}} y^{2n+1} = \frac{y}{1-y^2} +\text{arctanh}\,{y} \approx
\frac{i\pi}{2} -\frac{2}{3y^3},
\end{equation}
where the last approximation is valid for $y\gg 1$. Accordingly, we obtain that the imaginary part of the polarization operator close to the threshold is
\begin{equation}
\label{final result}
\chi''_{_V}(\omega) =-\frac{4N\gamma^2 }{ v \lambda_1^2} \left(\frac{\omega -\Delta}{\Delta}\right)^{3\gamma}.
\end{equation}
\begin{figure}
\resizebox{.35\textwidth}{!}{\includegraphics{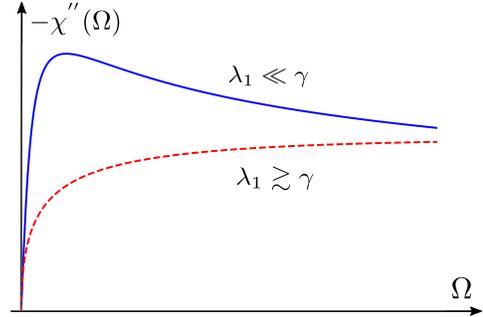}}
\caption{\label{fig3}[Color online] Absorption lineshape for different values of $V_1=4\pi v \lambda_1/N$. For small $V_1$ values  (solid blue line)  the suppression (depolarization) induced by it  overcomes the enhancement due to $V_0$ sufficiently close to the threshold. For larger $V_1$  values the depolarization effect dominates everywhere. }
\end{figure}

The obtained result means that the optical absorption (\ref{Joule heat}) close to the threshold $\omega=\Delta$ becomes suppressed due to the depolarization effect much more significantly than could be predicted based on the electrostatic mean field theory, $\propto (\omega -\Delta)^\gamma$. This can be viewed as a much stronger $E_i \propto (\omega -\Delta)^{2\gamma}E_0$ suppression of the electric field acting inside the wire, compared with Eq.~(\ref{electrostatics RPA series}).

The absorption lineshape is the result of an interplay between the symmetric $V_0$ and the dipolar $V_1$ interactions, see Fig.~\ref{fig3}. The result (\ref{final result}) holds sufficiently close to the threshold  no matter how small $V_1$ is: The latter is always relevant at $\Omega \to 0$. The magnitude of $V_1$ determines how close to the threshold the transition to the domain of Eq.~(\ref{final result}) happens. If $\lambda_1 \ll \gamma$, the interaction $V_1$ is not important far from the threshold, where the absorption is given by the imaginary part of Eq.~(\ref{polarization operator MS}), $\chi_{_V}'' =-N\Omega^{-\gamma}/v$, which increases with decreasing $\Omega$. At $\Omega^{2\gamma} \sim \lambda_1/\gamma$ the frequency dependence of the absorption crosses over to that of Eq.~(\ref{final result}) and drops sharply at the threshold. If, in contrast, the interaction $V_1$ is not very small, so that $\lambda_1 \gtrsim \gamma$, the expression (\ref{final result}) should be used everywhere near the threshold. This is what is expected to happen in metallic carbon nanotubes. The absorption given by Eqs.~(\ref{Joule heat}) and (\ref{final result}) should be testable in optical absorption measurements. Even though previous measurements of transverse absorption
\cite{LCT,LTT} do not provide sufficient resolution near the threshold, the modern advances in nanotube manufacturing  \cite{Haro} should make such measurement possible.

Finally, we note that although the depolarization effect exists in both semiconducting and metallic nanotubes, the physical mechanisms implicated in the two cases are different. As shown above, depolarization effect in metallic nanotubes is dominated by the shake-up of an infinite number of low-lying plasmon excitations. In contrast, in semiconducting nanotubes with no gapless states, where conventional excitonic effects can be expected to dominate, various {\it ab initio} approaches, such as the  Bethe-Salpeter method, should be used \cite{Spataru,Benedict}.

Discussions with O. Starykh, S. LeBohec, and D. Pesin are gratefully acknowledged. This work was supported
by DOE, Office of Basic Energy Sciences, Grant No.~DE-FG02-06ER46313.

\end{document}